\documentclass[prb, twocolumn, reprint, amsmath,amssymb]{revtex4-1}

\usepackage{graphicx,amssymb,amsmath,dsfont,epsfig}
\usepackage[latin1]{inputenc}

\newcommand{\wmk}{Wm$^{-1}$K$^{-1}$}

\begin{document}

\title{Atomistic simulations of heat transport in real-scale silicon nanowire devices}
\author{Ivan Duchemin}
\email{duchemin@mpip-mainz.mpg.de}
\affiliation{Laboratoire de Simulation Atomistique, SP2M, INAC, CEA, 38054 Grenoble, France}
\affiliation{Max Planck Institute for Polymer Research, Ackermannweg 10, 55128 Mainz, Germany}
\author{Davide Donadio}
\email{donadio@mpip-mainz.mpg.de}
\affiliation{Max Planck Institute for Polymer Research, Ackermannweg 10, 55128 Mainz, Germany}

\begin{abstract}
Utilizing atomistic lattice dynamics and scattering theory, we study thermal transport in nanodevices made of 10 nm thick silicon nanowires, from 10 to 100 nm long, sandwiched between two bulk reservoirs.
We find that thermal transport in devices differs significantly from that of suspended extended nanowires, due to phonon scattering at the contact interfaces.
We show that thermal conductance and the phonon transport regime can be tuned from ballistic to diffusive by varying the surface roughness of the nanowires and their length. In devices containing short crystalline wires phonon tunneling occurs and enhances the 
conductance beyond that of single contacts.
\end{abstract}


\maketitle

Silicon nanowires (SiNW) can be now routinely synthesized via different processes, such as VLS or etching, with very good control on diameter, length, alignment, growth direction and surface properties~\cite{Rurali:2010p2736}. Thus, they can be employed in electronic applications as parts of solid-state memories~\cite{Dong:2008p8} or field-effect transistors~\cite{Schmidt:2006,Appenzeller:2008p4174}.
Even though bulk Si is a poor thermoelectric material, SiNW have also been proposed as promising building blocks for thermoelectric applications~\cite{Dresselhaus:2007p3571}. Relatively high thermoelectric figures of merit were measured at room temperature, mostly due to a large reduction of lattice thermal conductivity~\cite{Boukai:2008p3187,Hochbaum:2008p3165}. There is indeed experimental evidence~\cite{Chen:2008p1755,Heron:2009p1725,Heron:2010p4178}, supported by theoretical studies~\cite{Murphy:2007p4175,Martin:2009p1718,Donadio:2009p1173,Sansoz:2011,Carrete:2011p4176} that the thermal conductivity of silicon at room temperature can be reduced by as much as two orders of magnitude by  nanostructuring and surface treatment~\cite{Ponomareva:2007p2068,Donadio:2010p3067}.

For any technological purpose it is essential to control heat transmission and dissipation in the device, carefully resolving the contribution from contact interfaces and/or supporting substrates~\cite{Pop:2010p4205,Moore:2011p4181}. Single crystalline SiNW with monolithic lead contacts were recently fabricated using electron beam lithography~\cite{Hippalgaonkar:2010p3802}. Their measured thermal conductivity (20 \wmk) resulted significantly higher than that of previous experiments on SiNW of similar diameter~\cite{Hochbaum:2008p3165} (2 \wmk), indicating that contacts play a major role in determining the overall thermal properties.


So far, atomistic modeling has been limited to extended periodic systems, whereas simulations of devices are commonly performed by mesoscale models~\cite{Shakouri:1998p4208,Ju:1999p4209,Lacroix:2006p4204}.
Here we utilize atomistic lattice dynamics to investigate thermal transport within SiNW devices made of 10 nm thick and up to 100 nm long wires connected coherently to two semi-infinite bulk reservoirs. The device thermal conductance is computed for wires of different lengths, with either crystalline or amorphized surfaces. We discuss how length, contacts and surface disorder determine different transport regimes. In particular, our calculations show that thermal transport in sub-micrometer devices is inherently different from that of extended wires, as it is controlled in large parts by contacts and finite size of the devices.


\begin{figure}[b]
\begin{center}
{\includegraphics[width=80mm]{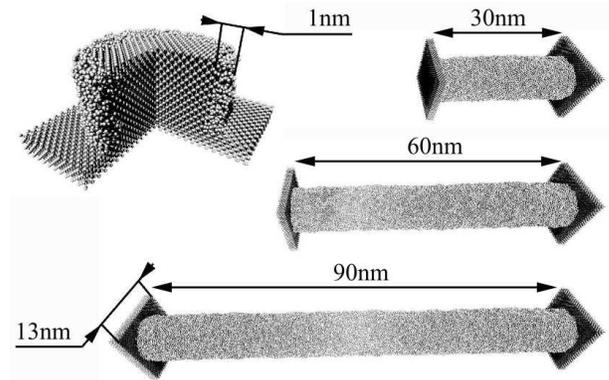}}
\end{center}
\caption{Representation of the core-shell silicon nanowire devices wires of different lengths, for which we have computed thermal conductance and heat transport properties. The silicon nanowires are connected coherently with semi-infinite bulk silicon thermal reservoirs. The core-shell nanowires shown here have a crystalline core of $\sim 8$ nm in diameter and a $\sim 1$ nm thick amorphized shell. Surface amorphization is obtained by heating the system slightly beyond its melting temperature (2500 K for the empirical model of silicon employed here) for 300 ps,  quenching to room temperature in 700 ps, and eventually optimizing the structure by steepest descent.}
\label{FigScheme}
\end{figure}

The wires are grown in the (100) crystallographic direction, have a nearly circular cross section and form ideal monolithic contacts with the reservoirs, preserving stacking and crystallographic orientation. Also, the surface of the crystalline SiNW is reconstructed in order to minimize the number of dangling bonds~\cite{Vo:2006p681}. Atomic positions of the wires are relaxed in a periodic cell at zero temperature and the cell length optimized so to obtain an unstrained system. The reservoirs are considered periodic in the $xy$ plane and semi-infinite in the $z$ plane. For both relaxation and heat transport calculations, the interactions between silicon atoms are modeled by means of the short-range empirical forcefield by Tersoff~\cite{JTersoff1989}.

We compute the thermal conductance of SiNW devices by harmonic lattice dynamics~\cite{Young:1989p3817,Fagas:1999p3575,Mingo:2003p110,Zhao:2005p024903,Wang:2006p3815}, employing the scattering matrix approach~\cite{Duchemin:2011p3936}.
This method permits the calculation of quantum thermal transport in open systems, yielding transmission spectra, thermal conductance and real-space representation of heat flux and non-equilibrium energy density. The latter is obtained by computing the local density of states (LDOS) and allows us to probe different transport regimes.
In spite of its assumptions and approximations, the most important of which is considering only elastic scattering, it has been recently demonstrated that this approach can consistently reproduce the Kapitza resistance of a silicon/germanium interface computed by molecular dynamics up to fairly large temperatures ($T\sim 500$ K)~\cite{Landry:2009p3275}. At higher temperatures anharmonic effects should be explicitly considered~\cite{Mingo:2006p3573}.



\begin{figure}[h]
\begin{center}
\includegraphics[width=80mm]{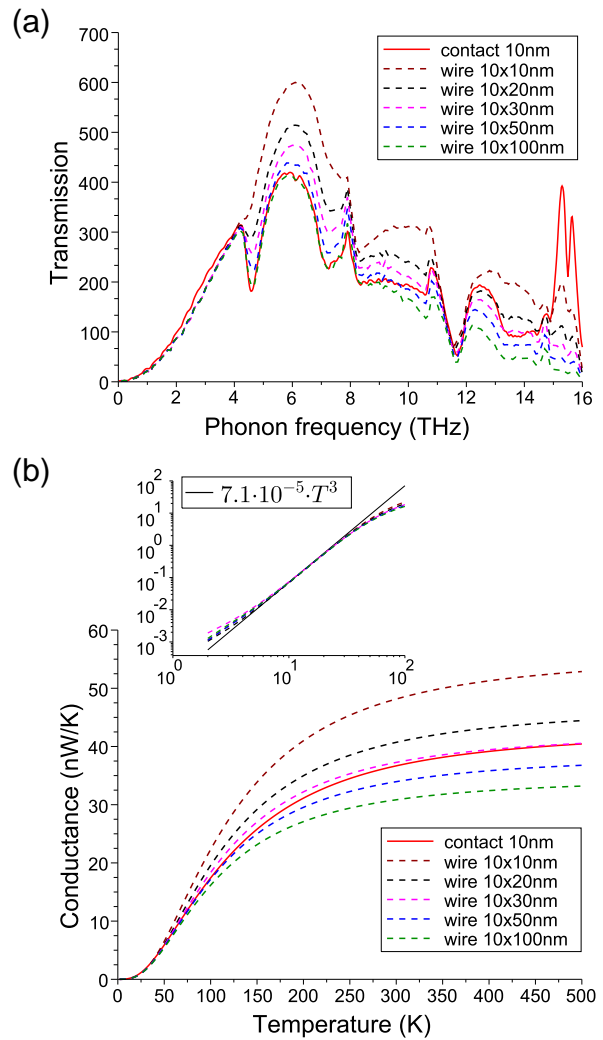}
\end{center}
\caption{Transmission spectra (a) and integrated thermal conductance (b) of devices containing crystalline silicon nanowires with 10 nm diameter and length varying between 10 and 100 nm (dashed lines), compared to the transmission spectrum and thermal conductance ($\sigma (T))$ of a single bulk/nanowire contact (solid line). The inset in panel (b) zooms in the low temperature region of $\sigma(T)$.
}
\label{Fig:transmission}
\end{figure}

\paragraph{Crystalline wires}
We first investigate the effect of varying the length of crystalline SiNW between two bulk reservoirs. Wire lengths of 10, 20, 30, 50 and 100 nm are considered: the largest system is made of $6\cdot 10^5$ atoms. 
The transmission spectra are shown in Fig.~\ref{Fig:transmission}a and compared to the transmission spectrum of the interface between semi-infinite bulk and the semi-infinite SiNW~\cite{Duchemin:2011p3936}.

\begin{figure}[h]
\begin{center}
\includegraphics[width=80mm]{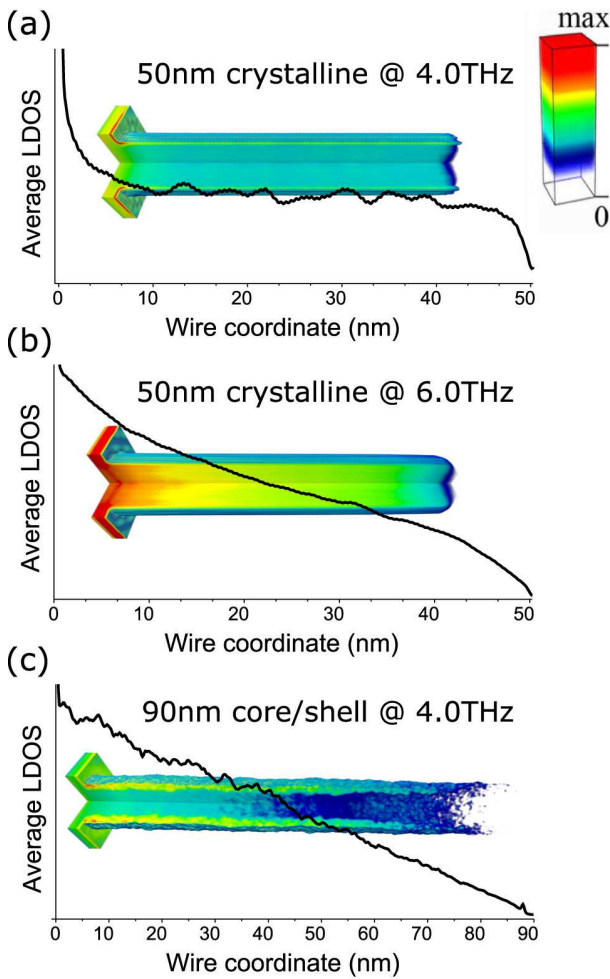}
\end{center}
\caption{Local density of states of crystalline wires (a-b) and core/shell wires (c). Each representation is associated to left hand incoming phonons contributions for the specified frequency.}
\label{Fig:ldos-all}
\end{figure}

We find that the transmission curves overlap for frequencies up to 4 THz regardless of the wire length (Fig.~\ref{Fig:transmission}), indicating that phonon transport at low frequency is ballistic. 
In this frequency range, there is no attenuation within the wire itself but only scattering at the bulk-wire interface. The device behaves as a complex Fabry-Perot cavity where the large number of channels average out the interference patterns. As a consequence, the transmission is independent on the length of the device, and so is the conductance, $\sigma(T)$, in the low temperature range. The ballistic regime below 4 THz was also probed computing the LDOS (Fig.~\ref{Fig:ldos-all}a), which indeed displays no attenuation within the wire. The LDOS decays sharply at the contacts and oscillates around a constant value. Those oscillations are reminiscent of the interference patterns formed within the device acting as a cavity.


\begin{figure}[h]
\label{Fig:channels}
\begin{center}
\includegraphics[width=80mm]{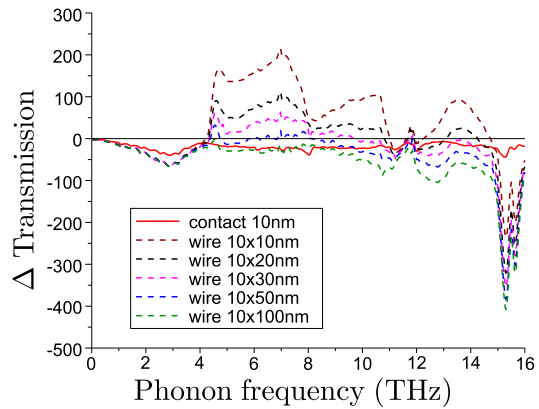}
\end{center}
\caption{Difference $\Delta$ between the transmission spectra of devices containing crystalline silicon nanowires and the transmission spectrum of an infinitely long silicon nanowire of the same type and diameter (i.e. its number of conduction channels). We consider devices made of wires of various length between 10 an 100 nm and diameter of 10 nm. For devices containing short wires ($< 50$) nm the difference is positive in a wide range of frequencies larger than 4 THz, indicating that in this range of frequency more modes are transmitted across the devices than the available conduction channels in the wire. Some bulk modes can tunnel through the device. The physical principle for phonon tunneling is analogous to the well known model of quantum tunneling across a finite barrier. The difference between the transmission spectrum of the contact interface and the number of channels of 10 nm wire is also reported for sake of comparison.}
\end{figure}
Surprisingly, the transmission of the devices becomes larger than that of the semi-infinite wire~\cite{Duchemin:2011p3936} for acoustic frequencies larger than 4 THz. We observe that the total transmission may even exceed the number of available channels in the wire (Fig. 4), indicating that a number of bulk states tunnels through the device. As a result, the total thermal conductance (Fig.~\ref{Fig:transmission}b) turns out larger than that of the single contact for wires shorter than 30 nm. The contribution from the tunneling states to $\sigma$ decreases when the length of the wires is increased, and vanishes for 100~nm and longer devices.
Plotting the local energy density allows direct visualization of the tunneling transport regime (Fig.~\ref{Fig:ldos-all}b). In contrast with the low frequency ballistic modes, the LDOS does not drop abruptly at the interface, but it penetrates the device region and decays slowly. 
A natural consequence of the combination of ballistic transport and phonon tunneling is that the thermal resistance of the two contact interfaces is not additive even for wires as long as 100 nm, contrary to what one would get for diffusive transport and contacts in series.

We also compare our computed thermal conductance with experimental measurements at very low temperature~\cite{Heron:2009p1725}, though they refer to much thicker nanowires ($\sim 100$ nm). In the temperature range between 5 and 20 K, $\sigma (T)$ is proportional to $T^3$, (inset of Fig.~\ref{Fig:transmission}b), indicating that dimensionality reduction has no significant effect on the low temperature trend of $\sigma$. The $T^3$ trend is the same as found for single contacts~\cite{Duchemin:2011p3936}, suggesting that at low temperature heat transport in these devices is determined by scattering at the contact interfaces. At variance with Ref.~\cite{Heron:2009p1725} we do not observe a temperature range where $\sigma (T) \propto T^2$ when SiNW are crystalline. This discrepancy is most likely due to the perfect crystallinity of our models, and different trends are observed when the surface of the SiNW is amorphized.

\begin{figure}[h]
\begin{center}
\includegraphics[width=70mm]{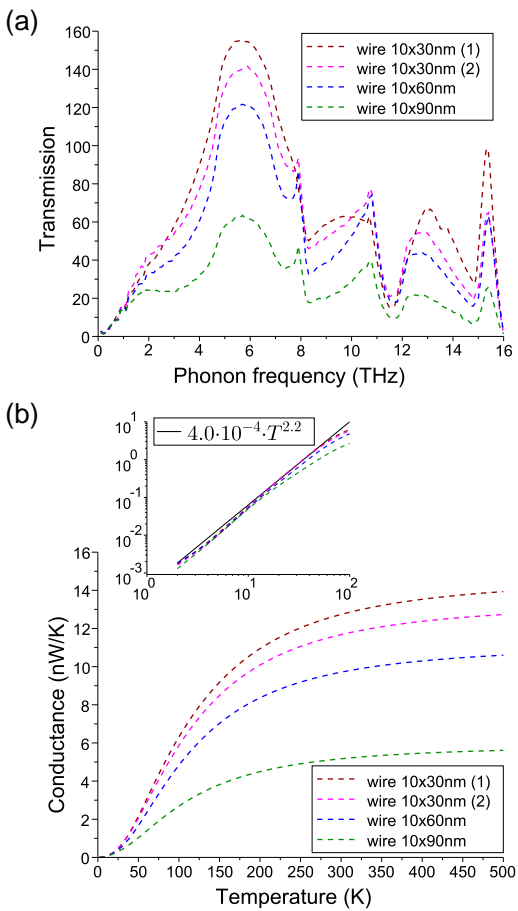}
\end{center}
\caption{Transmission spectra (a) and integrated thermal conductance (b) of crystalline core-amorphous shell silicon nanowires with 10 nm diameter and lengths varying between 30 and 90 nm. The inset in panel (b) zooms in the low temperature region of $\sigma(T)$.}
\label{Fig:transmission_rough}
\end{figure}

\paragraph{Core-shell wires}
We now discuss  thermal transport in devices with crystalline core/amorphous shell silicon nanowires (CS-SiNW). To obtain CS-SiNW, crystalline SiNW are heated beyond the melting temperature of the bulk for a short time (300 ps), quenched to room temperature in 700 ps and optimized using the steepest descent algorithm. This procedure produces a $\sim 1$ nm thick amorphous layer.
Four models have been designed for this purpose: two with 30~nm long wires, one with 60~nm and one with 90~nm long wires. The surface disorder in the first 30~nm sample is limited to about two surface layers, whereas the amorphous shell of the remaining 30~nm, 60~nm and 90~nm samples  is about 1~nm thick (Fig. 1).
The transmission spectra and the thermal conductance of these devices are shown in Fig.~\ref{Fig:transmission_rough}.
With respect to crystalline devices with wires of similar length, we observe that shell amorphization leads to a reduction of the device thermal conductance by a factor between 3 and 6 at room temperature. The larger reduction occurring in the longer device.  
For sub-micrometer wire devices with diameter above 10 nm, surface amorphization does not have an effect on thermal transport as dramatic as that observed in thinner SiNW~\cite{Donadio:2009p1173}. This indicates that when large thermal resistance is experimentally observed for SiNW thicker than 10 nm, surface disorder has to be supplemented by other sources of phonon scattering, such as extended defects or poor contact interfaces.

The inset in Fig.~\ref{Fig:transmission_rough}b shows that at low temperature $\sigma(T)\propto T^{2.2}$. This indicates that elastic scattering at the crystalline/amorphous interface can be the cause for the  $T^2$ trends of thermal conductance observed experimentally in the temperature range between 1 and 10 K~\cite{Heron:2009p1725}. Dimensionality reduction plays the indirect role of enhancing surface effects, due to the higher surface-to-volume ration in nanostructures. At variance with experiments, the $T^2$ range for $\sigma(T)$ extends to 50 K: this discrepancy may be due either to size differences between models and experiments (experimental SiNW have much larger cross section), or to neglecting of anharmonic scattering processes in our lattice dynamics approach.

The calculation of LDOS in the CS-SiNW devices shows that phonons with frequency up to 1~THz are transmitted almost ballistically along the wire. The transmission coefficient is not sensitive to either the wire length or the thickness of the amorphous layer, and is comparable to the one of the crystalline devices. (Fig.~\ref{Fig:transmission_rough}a).
At higher frequencies, the phonons undergo multiple scattering from the rough crystalline-amorphous interface and the LDOS decays linearly along the wire, as shown in Fig.~\ref{Fig:ldos-all}c. The transmission coefficient drops accordingly, becoming significantly lower than the one of the crystalline device.
This behavior is characteristic of the diffusive transport regime. We also observe that the LDOS is larger in the amorphized shell of the wires, which seems to indicate that the shell is warmer than the core in non-equilibrium conditions.
For phonon frequencies up to 2 THz, the LDOS and heat flux retain similar features as for crystalline wires. At higher frequencies, the heat flux is concentrated within the core region of the wire and presents irregularities, confirming the effect of the shell amorphization on the wire conduction properties.

\begin{figure}[h]
\begin{center}
\includegraphics[width=70mm]{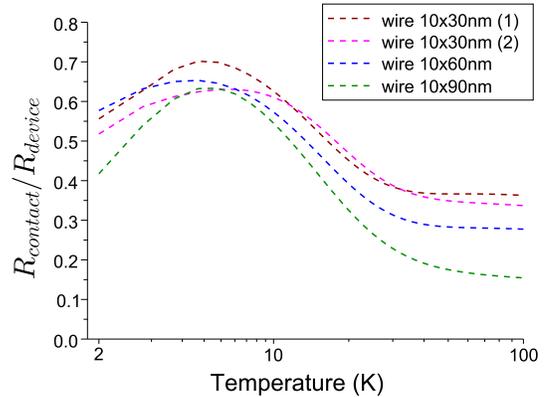}
\end{center}
\caption{Ratio between the resistance of a single crystalline contact and the total resistance of the devices with core-shell nanowires 30, 60 and 90 nm long.}
\label{Fig:contacts}
\end{figure}
Assuming that the contact resistance does not vary significantly from crystalline to core-shell wires, we use the values obtained in a previous work~\cite{Duchemin:2011p3936}, and compute the ratio between a single contact resistance and the total resistances of the core-shell SiNW devices (Fig.~\ref{Fig:contacts}).
The relative importance of contact thermal resistance depends on the temperature and on the length of the wires. Contact resistance prevails at low temperatures, reaching a maximum over 60$\%$ of the total device resistance below 10 K, and becomes less important at higher temperatures.
In the low temperature regime only low frequency modes are populated, which conduct ballistically through the wires: contacts are the only source of scattering.
For $T$ larger than 100 K the curves in Fig.~\ref{Fig:contacts} reach a plateau, which value depends on the length of the wires. The relative effect of contact resistance at $T>100$ K decreases from $\sim 35\%$ for the 30 nm wire device to $\sim 15\%$ for the 90 nm wire device.
For longer wires the relative weight of contact resistance would decrease further, as the wire resistance would increase linearly with the length, yet our estimate is a lower boundary, since we consider ideal monolithic contacts. For instance, if we assume that we can treat the 90 nm wire and the contacts as thermal resistors in series, and evaluate each contribution as additive, then the contacts contribute  about 30$\%$ of the total thermal resistance of the device at room temperature.


In summary, our calculations prove that:
i) Thermal transport is ballistic in devices containing crystalline nanowires, and scattering of acoustic phonons occurs mostly at the contact interfaces.
However, when SiNW are shorter than 30 nm the thermal conductance may be larger than the contact conductance, due to tunneling of bulk vibrational modes through the SiNW.
ii) Amorphization of a surface layer can reduce the thermal conductance of these type of devices by at most six times at room temperature, as it is ineffective in scattering low frequency acoustic modes. The introduction of long-range disorder or nano-structuring is probably essential to further decrease the low frequencies contribution to thermal conductance.
iii) The contribution of contact interface scattering to the thermal resistance is significant even in those systems where there is strong surface scattering and heat transport is mostly diffusive.
iv) In CS-SiNW devices the thermal conductance at low temperature is proportional to $T^2$, in remarkably good agreement with experiments~\cite{Heron:2009p1725}. We note that the surface scattering term is sufficient to account for this unusual trend of $\sigma(T)$.

\acknowledgements{
Calculations were performed at the Rechenzentrum Garching of the Max Planck Society.
D.D. acknowledges funding from the Max Planck Society under the MPRG program.
We thank O. Bourgeois, G. Galli, L.~F. Pereira and D. Andrienko for useful discussions and critical reading of the manuscript.
}


\begin{thebibliography}{33}%
\makeatletter
\providecommand \@ifxundefined [1]{%
 \@ifx{#1\undefined}
}%
\providecommand \@ifnum [1]{%
 \ifnum #1\expandafter \@firstoftwo
 \else \expandafter \@secondoftwo
 \fi
}%
\providecommand \@ifx [1]{%
 \ifx #1\expandafter \@firstoftwo
 \else \expandafter \@secondoftwo
 \fi
}%
\providecommand \natexlab [1]{#1}%
\providecommand \enquote  [1]{``#1''}%
\providecommand \bibnamefont  [1]{#1}%
\providecommand \bibfnamefont [1]{#1}%
\providecommand \citenamefont [1]{#1}%
\providecommand \href@noop [0]{\@secondoftwo}%
\providecommand \href [0]{\begingroup \@sanitize@url \@href}%
\providecommand \@href[1]{\@@startlink{#1}\@@href}%
\providecommand \@@href[1]{\endgroup#1\@@endlink}%
\providecommand \@sanitize@url [0]{\catcode `\\12\catcode `\$12\catcode
  `\&12\catcode `\#12\catcode `\^12\catcode `\_12\catcode `\%12\relax}%
\providecommand \@@startlink[1]{}%
\providecommand \@@endlink[0]{}%
\providecommand \url  [0]{\begingroup\@sanitize@url \@url }%
\providecommand \@url [1]{\endgroup\@href {#1}{\urlprefix }}%
\providecommand \urlprefix  [0]{URL }%
\providecommand \Eprint [0]{\href }%
\providecommand \doibase [0]{http://dx.doi.org/}%
\providecommand \selectlanguage [0]{\@gobble}%
\providecommand \bibinfo  [0]{\@secondoftwo}%
\providecommand \bibfield  [0]{\@secondoftwo}%
\providecommand \translation [1]{[#1]}%
\providecommand \BibitemOpen [0]{}%
\providecommand \bibitemStop [0]{}%
\providecommand \bibitemNoStop [0]{.\EOS\space}%
\providecommand \EOS [0]{\spacefactor3000\relax}%
\providecommand \BibitemShut  [1]{\csname bibitem#1\endcsname}%
\let\auto@bib@innerbib\@empty
\bibitem [{\citenamefont {Rurali}(2010)}]{Rurali:2010p2736}%
  \BibitemOpen
  \bibfield  {author} {\bibinfo {author} {\bibfnamefont {R.}~\bibnamefont
  {Rurali}},\ }\href@noop {} {\bibfield  {journal} {\bibinfo  {journal}
  {Rev. Mod. Phys.}\ }\textbf {\bibinfo {volume} {82}},\ \bibinfo
  {pages} {427} (\bibinfo {year} {2010})}\BibitemShut {NoStop}%
\bibitem [{\citenamefont {Dong}\ \emph {et~al.}(2008)\citenamefont {Dong},
  \citenamefont {Yu}, \citenamefont {Mcalpine}, \citenamefont {Lu},\ and\
  \citenamefont {Lieber}}]{Dong:2008p8}%
  \BibitemOpen
  \bibfield  {author} {\bibinfo {author} {\bibfnamefont {Y.}~\bibnamefont
  {Dong}}, \bibinfo {author} {\bibfnamefont {G.}~\bibnamefont {Yu}}, \bibinfo
  {author} {\bibfnamefont {M.~C.}\ \bibnamefont {Mcalpine}}, \bibinfo {author}
  {\bibfnamefont {W.}~\bibnamefont {Lu}}, \ and\ \bibinfo {author}
  {\bibfnamefont {C.~M.}\ \bibnamefont {Lieber}},\ }\href@noop {} {\bibfield
  {journal} {\bibinfo  {journal} {Nano Lett.}\ }\textbf {\bibinfo {volume}
  {8}},\ \bibinfo {pages} {386} (\bibinfo {year} {2008})}\BibitemShut {NoStop}%
\bibitem [{\citenamefont {Schmidt}\ \emph {et~al.}(2006)\citenamefont
  {Schmidt}, \citenamefont {Riel}, \citenamefont {Senz}, \citenamefont {Karg},
  \citenamefont {Riess},\ and\ \citenamefont {Goesele}}]{Schmidt:2006}%
  \BibitemOpen
  \bibfield  {author} {\bibinfo {author} {\bibfnamefont {V.}~\bibnamefont
  {Schmidt}}, \bibinfo {author} {\bibfnamefont {H.}~\bibnamefont {Riel}},
  \bibinfo {author} {\bibfnamefont {S.}~\bibnamefont {Senz}}, \bibinfo {author}
  {\bibfnamefont {S.}~\bibnamefont {Karg}}, \bibinfo {author} {\bibfnamefont
  {W.}~\bibnamefont {Riess}}, \ and\ \bibinfo {author} {\bibfnamefont
  {U.}~\bibnamefont {Goesele}},\ }\href@noop {} {\bibfield  {journal} {\bibinfo
   {journal} {Small}\ }\textbf {\bibinfo {volume} {2}},\ \bibinfo {pages} {85}
  (\bibinfo {year} {2006})}\BibitemShut {NoStop}%
\bibitem [{\citenamefont {Appenzeller}\ \emph {et~al.}(2008)\citenamefont
  {Appenzeller}, \citenamefont {Knoch}, \citenamefont {Bjoerk}, \citenamefont
  {Riel}, \citenamefont {Schmid},\ and\ \citenamefont
  {Riess}}]{Appenzeller:2008p4174}%
  \BibitemOpen
  \bibfield  {author} {\bibinfo {author} {\bibfnamefont {J.}~\bibnamefont
  {Appenzeller}}, \bibinfo {author} {\bibfnamefont {J.}~\bibnamefont {Knoch}},
  \bibinfo {author} {\bibfnamefont {M.~I.}\ \bibnamefont {Bjoerk}}, \bibinfo
  {author} {\bibfnamefont {H.}~\bibnamefont {Riel}}, \bibinfo {author}
  {\bibfnamefont {H.}~\bibnamefont {Schmid}}, \ and\ \bibinfo {author}
  {\bibfnamefont {W.}~\bibnamefont {Riess}},\ }\href@noop {} {\bibfield
  {journal} {\bibinfo  {journal} {IEEE Trans. Electron. Dev.}\ }\textbf {\bibinfo
  {volume} {55}},\ \bibinfo {pages} {2827} (\bibinfo {year}
  {2008})}\BibitemShut {NoStop}%
\bibitem [{\citenamefont {Dresselhaus}\ \emph {et~al.}(2007)\citenamefont
  {Dresselhaus}, \citenamefont {Chen}, \citenamefont {Tang}, \citenamefont
  {Yang}, \citenamefont {Lee}, \citenamefont {Wang}, \citenamefont {Ren},
  \citenamefont {Fleurial},\ and\ \citenamefont
  {Gogna}}]{Dresselhaus:2007p3571}%
  \BibitemOpen
  \bibfield  {author} {\bibinfo {author} {\bibfnamefont {M.~S.}\ \bibnamefont
  {Dresselhaus}}, \bibinfo {author} {\bibfnamefont {G.}~\bibnamefont {Chen}},
  \bibinfo {author} {\bibfnamefont {M.~Y.}\ \bibnamefont {Tang}}, \bibinfo
  {author} {\bibfnamefont {R.}~\bibnamefont {Yang}}, \bibinfo {author}
  {\bibfnamefont {H.}~\bibnamefont {Lee}}, \bibinfo {author} {\bibfnamefont
  {D.}~\bibnamefont {Wang}}, \bibinfo {author} {\bibfnamefont {Z.}~\bibnamefont
  {Ren}}, \bibinfo {author} {\bibfnamefont {J.-P.}\ \bibnamefont {Fleurial}}, \
  and\ \bibinfo {author} {\bibfnamefont {P.}~\bibnamefont {Gogna}},\
  }\href@noop {} {\bibfield  {journal} {\bibinfo  {journal} {Adv. Mater.}\
  }\textbf {\bibinfo {volume} {19}},\ \bibinfo {pages} {1043} (\bibinfo {year}
  {2007})}\BibitemShut {NoStop}%
\bibitem [{\citenamefont {Boukai}\ \emph {et~al.}(2008)\citenamefont {Boukai},
  \citenamefont {Bunimovich}, \citenamefont {Tahir-Kheli}, \citenamefont {Yu},
  \citenamefont {Goddard},\ and\ \citenamefont {Heath}}]{Boukai:2008p3187}%
  \BibitemOpen
  \bibfield  {author} {\bibinfo {author} {\bibfnamefont {A.~I.}\ \bibnamefont
  {Boukai}}, \bibinfo {author} {\bibfnamefont {Y.}~\bibnamefont {Bunimovich}},
  \bibinfo {author} {\bibfnamefont {J.}~\bibnamefont {Tahir-Kheli}}, \bibinfo
  {author} {\bibfnamefont {J.-K.}\ \bibnamefont {Yu}}, \bibinfo {author}
  {\bibfnamefont {W.~A.}\ \bibnamefont {Goddard}}, \ and\ \bibinfo {author}
  {\bibfnamefont {J.~R.}\ \bibnamefont {Heath}},\ }\href@noop {} {\bibfield
  {journal} {\bibinfo  {journal} {Nature}\ }\textbf {\bibinfo {volume} {451}},\
  \bibinfo {pages} {168} (\bibinfo {year} {2008})}\BibitemShut {NoStop}%
\bibitem [{\citenamefont {Hochbaum}\ \emph {et~al.}(2008)\citenamefont
  {Hochbaum}, \citenamefont {Chen}, \citenamefont {Delgado}, \citenamefont
  {Liang}, \citenamefont {Garnett}, \citenamefont {Najarian}, \citenamefont
  {Majumdar},\ and\ \citenamefont {Yang}}]{Hochbaum:2008p3165}%
  \BibitemOpen
  \bibfield  {author} {\bibinfo {author} {\bibfnamefont {A.~I.}\ \bibnamefont
  {Hochbaum}}, \bibinfo {author} {\bibfnamefont {R.}~\bibnamefont {Chen}},
  \bibinfo {author} {\bibfnamefont {R.~D.}\ \bibnamefont {Delgado}}, \bibinfo
  {author} {\bibfnamefont {W.}~\bibnamefont {Liang}}, \bibinfo {author}
  {\bibfnamefont {E.~C.}\ \bibnamefont {Garnett}}, \bibinfo {author}
  {\bibfnamefont {M.}~\bibnamefont {Najarian}}, \bibinfo {author}
  {\bibfnamefont {A.}~\bibnamefont {Majumdar}}, \ and\ \bibinfo {author}
  {\bibfnamefont {P.}~\bibnamefont {Yang}},\ }\href@noop {} {\bibfield
  {journal} {\bibinfo  {journal} {Nature}\ }\textbf {\bibinfo {volume} {451}},\
  \bibinfo {pages} {163} (\bibinfo {year} {2008})}\BibitemShut {NoStop}%
\bibitem [{\citenamefont {Chen}\ \emph {et~al.}(2008)\citenamefont {Chen},
  \citenamefont {Hochbaum}, \citenamefont {Murphy}, \citenamefont {Moore},
  \citenamefont {Yang},\ and\ \citenamefont {Majumdar}}]{Chen:2008p1755}%
  \BibitemOpen
  \bibfield  {author} {\bibinfo {author} {\bibfnamefont {R.}~\bibnamefont
  {Chen}}, \bibinfo {author} {\bibfnamefont {A.~I.}\ \bibnamefont {Hochbaum}},
  \bibinfo {author} {\bibfnamefont {P.}~\bibnamefont {Murphy}}, \bibinfo
  {author} {\bibfnamefont {J.}~\bibnamefont {Moore}}, \bibinfo {author}
  {\bibfnamefont {P.}~\bibnamefont {Yang}}, \ and\ \bibinfo {author}
  {\bibfnamefont {A.}~\bibnamefont {Majumdar}},\ }\href@noop {} {\bibfield
  {journal} {\bibinfo  {journal} {Phys. Rev. Lett.}\ }\textbf {\bibinfo
  {volume} {101}},\ \bibinfo {pages} {105501} (\bibinfo {year}
  {2008})}\BibitemShut {NoStop}%
\bibitem [{\citenamefont {Heron}\ \emph {et~al.}(2009)\citenamefont {Heron},
  \citenamefont {Fournier}, \citenamefont {Mingo},\ and\ \citenamefont
  {Bourgeois}}]{Heron:2009p1725}%
  \BibitemOpen
  \bibfield  {author} {\bibinfo {author} {\bibfnamefont {J.~S.}\ \bibnamefont
  {Heron}}, \bibinfo {author} {\bibfnamefont {T.}~\bibnamefont {Fournier}},
  \bibinfo {author} {\bibfnamefont {N.}~\bibnamefont {Mingo}}, \ and\ \bibinfo
  {author} {\bibfnamefont {O.}~\bibnamefont {Bourgeois}},\ }\href@noop {}
  {\bibfield  {journal} {\bibinfo  {journal} {Nano Lett.}\ }\textbf {\bibinfo
  {volume} {9}},\ \bibinfo {pages} {1861} (\bibinfo {year} {2009})}\BibitemShut
  {NoStop}%
\bibitem [{\citenamefont {Heron}\ \emph {et~al.}(2010)\citenamefont {Heron},
  \citenamefont {Bera}, \citenamefont {Fournier}, \citenamefont {Mingo},\ and\
  \citenamefont {Bourgeois}}]{Heron:2010p4178}%
  \BibitemOpen
  \bibfield  {author} {\bibinfo {author} {\bibfnamefont {J.-S.}\ \bibnamefont
  {Heron}}, \bibinfo {author} {\bibfnamefont {C.}~\bibnamefont {Bera}},
  \bibinfo {author} {\bibfnamefont {T.}~\bibnamefont {Fournier}}, \bibinfo
  {author} {\bibfnamefont {N.}~\bibnamefont {Mingo}}, \ and\ \bibinfo {author}
  {\bibfnamefont {O.}~\bibnamefont {Bourgeois}},\ }\href@noop {} {\bibfield
  {journal} {\bibinfo  {journal} {Phys. Rev. B}\ }\textbf {\bibinfo {volume}
  {82}},\ \bibinfo {pages} {155458} (\bibinfo {year} {2010})}\BibitemShut
  {NoStop}%
\bibitem [{\citenamefont {Murphy}\ and\ \citenamefont
  {Moore}(2007)}]{Murphy:2007p4175}%
  \BibitemOpen
  \bibfield  {author} {\bibinfo {author} {\bibfnamefont {P.~G.}\ \bibnamefont
  {Murphy}}\ and\ \bibinfo {author} {\bibfnamefont {J.~E.}\ \bibnamefont
  {Moore}},\ }\href@noop {} {\bibfield  {journal} {\bibinfo  {journal} {Phys.
  Rev. B}\ }\textbf {\bibinfo {volume} {76}},\ \bibinfo {pages} {155313}
  (\bibinfo {year} {2007})}\BibitemShut {NoStop}%
\bibitem [{\citenamefont {Martin}\ \emph {et~al.}(2009)\citenamefont {Martin},
  \citenamefont {Aksamija}, \citenamefont {Pop},\ and\ \citenamefont
  {Ravaioli}}]{Martin:2009p1718}%
  \BibitemOpen
  \bibfield  {author} {\bibinfo {author} {\bibfnamefont {P.}~\bibnamefont
  {Martin}}, \bibinfo {author} {\bibfnamefont {Z.}~\bibnamefont {Aksamija}},
  \bibinfo {author} {\bibfnamefont {E.}~\bibnamefont {Pop}}, \ and\ \bibinfo
  {author} {\bibfnamefont {U.}~\bibnamefont {Ravaioli}},\ }\href@noop {}
  {\bibfield  {journal} {\bibinfo  {journal} {Phys. Rev. Lett.}\ }\textbf
  {\bibinfo {volume} {102}},\ \bibinfo {pages} {125503} (\bibinfo {year}
  {2009})}\BibitemShut {NoStop}%
\bibitem [{\citenamefont {Donadio}\ and\ \citenamefont
  {Galli}(2009)}]{Donadio:2009p1173}%
  \BibitemOpen
  \bibfield  {author} {\bibinfo {author} {\bibfnamefont {D.}~\bibnamefont
  {Donadio}}\ and\ \bibinfo {author} {\bibfnamefont {G.}~\bibnamefont
  {Galli}},\ }\href@noop {} {\bibfield  {journal} {\bibinfo  {journal} {Phys.
  Rev. Lett.}\ }\textbf {\bibinfo {volume} {102}},\ \bibinfo {pages} {195901}
  (\bibinfo {year} {2009})}\BibitemShut {NoStop}%
\bibitem [{\citenamefont {Sansoz}(2011)}]{Sansoz:2011}%
  \BibitemOpen
  \bibfield  {author} {\bibinfo {author} {\bibfnamefont {F.}~\bibnamefont
  {Sansoz}},\ }\href@noop {} {\bibfield  {journal} {\bibinfo  {journal} {Nano
  Lett.}\ }\textbf {\bibinfo {volume} {11}},\ \bibinfo {pages} {5378}
  (\bibinfo {year} {2011})}\BibitemShut {NoStop}%
\bibitem [{\citenamefont {Carrete}\ \emph {et~al.}(2011)\citenamefont
  {Carrete}, \citenamefont {Gallego}, \citenamefont {Varela},\ and\
  \citenamefont {Mingo}}]{Carrete:2011p4176}%
  \BibitemOpen
  \bibfield  {author} {\bibinfo {author} {\bibfnamefont {J.}~\bibnamefont
  {Carrete}}, \bibinfo {author} {\bibfnamefont {L.~J.}\ \bibnamefont
  {Gallego}}, \bibinfo {author} {\bibfnamefont {L.~M.}\ \bibnamefont {Varela}},
  \ and\ \bibinfo {author} {\bibfnamefont {N.}~\bibnamefont {Mingo}},\
  }\href@noop {} {\bibfield  {journal} {\bibinfo  {journal} {Phys. Rev. B}\
  }\textbf {\bibinfo {volume} {84}},\ \bibinfo {pages} {075403} (\bibinfo
  {year} {2011})}\BibitemShut {NoStop}%
\bibitem [{\citenamefont {Ponomareva}, \citenamefont {Srivastava},\ and\
  \citenamefont {Menon}(2007)}]{Ponomareva:2007p2068}%
  \BibitemOpen
  \bibfield  {author} {\bibinfo {author} {\bibfnamefont {I.}~\bibnamefont
  {Ponomareva}}, \bibinfo {author} {\bibfnamefont {D.}~\bibnamefont
  {Srivastava}}, \ and\ \bibinfo {author} {\bibfnamefont {M.}~\bibnamefont
  {Menon}},\ }\href@noop {} {\bibfield  {journal} {\bibinfo  {journal} {Nano
  Lett.}\ }\textbf {\bibinfo {volume} {7}},\ \bibinfo {pages} {1155} (\bibinfo
  {year} {2007})}\BibitemShut {NoStop}%
\bibitem [{\citenamefont {Donadio}\ and\ \citenamefont
  {Galli}(2010)}]{Donadio:2010p3067}%
  \BibitemOpen
  \bibfield  {author} {\bibinfo {author} {\bibfnamefont {D.}~\bibnamefont
  {Donadio}}\ and\ \bibinfo {author} {\bibfnamefont {G.}~\bibnamefont
  {Galli}},\ }\href@noop {} {\bibfield  {journal} {\bibinfo  {journal} {Nano
  Lett.}\ }\textbf {\bibinfo {volume} {10}},\ \bibinfo {pages} {847} (\bibinfo
  {year} {2010})}\BibitemShut {NoStop}%
\bibitem [{\citenamefont {Pop}(2010)}]{Pop:2010p4205}%
  \BibitemOpen
  \bibfield  {author} {\bibinfo {author} {\bibfnamefont {E.}~\bibnamefont
  {Pop}},\ }\href@noop {} {\bibfield  {journal} {\bibinfo  {journal} {Nano
  Res.}\ }\textbf {\bibinfo {volume} {3}},\ \bibinfo {pages} {147} (\bibinfo
  {year} {2010})}\BibitemShut {NoStop}%
\bibitem [{\citenamefont {Moore}\ and\ \citenamefont
  {Shi}(2011)}]{Moore:2011p4181}%
  \BibitemOpen
  \bibfield  {author} {\bibinfo {author} {\bibfnamefont {A.~L.}\ \bibnamefont
  {Moore}}\ and\ \bibinfo {author} {\bibfnamefont {L.}~\bibnamefont {Shi}},\
  }\href@noop {} {\bibfield  {journal} {\bibinfo  {journal} {Meas Sci Technol}\
  }\textbf {\bibinfo {volume} {22}},\ \bibinfo {pages} {015103} (\bibinfo
  {year} {2011})}\BibitemShut {NoStop}%
\bibitem [{\citenamefont {Hippalgaonkar}\ \emph {et~al.}(2010)\citenamefont
  {Hippalgaonkar}, \citenamefont {Huang}, \citenamefont {Chen}, \citenamefont
  {Sawyer}, \citenamefont {Ercius},\ and\ \citenamefont
  {Majumdar}}]{Hippalgaonkar:2010p3802}%
  \BibitemOpen
  \bibfield  {author} {\bibinfo {author} {\bibfnamefont {K.}~\bibnamefont
  {Hippalgaonkar}}, \bibinfo {author} {\bibfnamefont {B.}~\bibnamefont
  {Huang}}, \bibinfo {author} {\bibfnamefont {R.}~\bibnamefont {Chen}},
  \bibinfo {author} {\bibfnamefont {K.}~\bibnamefont {Sawyer}}, \bibinfo
  {author} {\bibfnamefont {P.}~\bibnamefont {Ercius}}, \ and\ \bibinfo {author}
  {\bibfnamefont {A.}~\bibnamefont {Majumdar}},\ }\href@noop {} {\bibfield
  {journal} {\bibinfo  {journal} {Nano Lett.}\ }\textbf {\bibinfo {volume}
  {10}},\ \bibinfo {pages} {4341} (\bibinfo {year} {2010})}\BibitemShut
  {NoStop}%
\bibitem [{\citenamefont {Shakouri}\ \emph {et~al.}(1998)\citenamefont
  {Shakouri}, \citenamefont {Lee}, \citenamefont {Smith}, \citenamefont
  {Narayanamurti},\ and\ \citenamefont {Bowers}}]{Shakouri:1998p4208}%
  \BibitemOpen
  \bibfield  {author} {\bibinfo {author} {\bibfnamefont {A.}~\bibnamefont
  {Shakouri}}, \bibinfo {author} {\bibfnamefont {E.}~\bibnamefont {Lee}},
  \bibinfo {author} {\bibfnamefont {D.}~\bibnamefont {Smith}}, \bibinfo
  {author} {\bibfnamefont {V.}~\bibnamefont {Narayanamurti}}, \ and\ \bibinfo
  {author} {\bibfnamefont {J.}~\bibnamefont {Bowers}},\ }\href@noop {}
  {\bibfield  {journal} {\bibinfo  {journal} {Microscale Therm Eng}\ }\textbf
  {\bibinfo {volume} {2}},\ \bibinfo {pages} {37} (\bibinfo {year}
  {1998})}\BibitemShut {NoStop}%
\bibitem [{\citenamefont {Ju}\ and\ \citenamefont
  {Goodson}(1999)}]{Ju:1999p4209}%
  \BibitemOpen
  \bibfield  {author} {\bibinfo {author} {\bibfnamefont {Y.}~\bibnamefont
  {Ju}}\ and\ \bibinfo {author} {\bibfnamefont {K.}~\bibnamefont {Goodson}},\
  }\href@noop {} {\bibfield  {journal} {\bibinfo  {journal} {Appl. Phys. Lett.}\
  }\textbf {\bibinfo {volume} {74}},\ \bibinfo {pages} {3005} (\bibinfo {year}
  {1999})}\BibitemShut {NoStop}%
\bibitem [{\citenamefont {Lacroix}\ \emph {et~al.}(2006)\citenamefont
  {Lacroix}, \citenamefont {Joulain}, \citenamefont {Terris},\ and\
  \citenamefont {Lemonnier}}]{Lacroix:2006p4204}%
  \BibitemOpen
  \bibfield  {author} {\bibinfo {author} {\bibfnamefont {D.}~\bibnamefont
  {Lacroix}}, \bibinfo {author} {\bibfnamefont {K.}~\bibnamefont {Joulain}},
  \bibinfo {author} {\bibfnamefont {D.}~\bibnamefont {Terris}}, \ and\ \bibinfo
  {author} {\bibfnamefont {D.}~\bibnamefont {Lemonnier}},\ }\href@noop {}
  {\bibfield  {journal} {\bibinfo  {journal} {Appl. Phys. Lett.}\ }\textbf
  {\bibinfo {volume} {89}},\ \bibinfo {pages} {103104} (\bibinfo {year}
  {2006})}\BibitemShut {NoStop}%
\bibitem [{\citenamefont {Vo}, \citenamefont {Williamson},\ and\ \citenamefont
  {Galli}(2006)}]{Vo:2006p681}%
  \BibitemOpen
  \bibfield  {author} {\bibinfo {author} {\bibfnamefont {T.}~\bibnamefont
  {Vo}}, \bibinfo {author} {\bibfnamefont {A.~J.}\ \bibnamefont {Williamson}},
  \ and\ \bibinfo {author} {\bibfnamefont {G.}~\bibnamefont {Galli}},\
  }\href@noop {} {\bibfield  {journal} {\bibinfo  {journal} {Phys. Rev. B}\
  }\textbf {\bibinfo {volume} {74}},\ \bibinfo {pages} {045116} (\bibinfo
  {year} {2006})}\BibitemShut {NoStop}%
\bibitem [{\citenamefont {Tersoff}(1989)}]{JTersoff1989}%
  \BibitemOpen
  \bibfield  {author} {\bibinfo {author} {\bibfnamefont {J.}~\bibnamefont
  {Tersoff}},\ }\href@noop {} {\bibfield  {journal} {\bibinfo  {journal} {Phys.
  Rev. B}\ }\textbf {\bibinfo {volume} {39}},\ \bibinfo {pages} {5566}
  (\bibinfo {year} {1989})}\BibitemShut {NoStop}%
\bibitem [{\citenamefont {Young}\ and\ \citenamefont
  {Maris}(1989)}]{Young:1989p3817}%
  \BibitemOpen
  \bibfield  {author} {\bibinfo {author} {\bibfnamefont {D.~A.}\ \bibnamefont
  {Young}}\ and\ \bibinfo {author} {\bibfnamefont {H.~J.}\ \bibnamefont
  {Maris}},\ }\href@noop {} {\bibfield  {journal} {\bibinfo  {journal} {Phys.
  Rev. B}\ }\textbf {\bibinfo {volume} {40}},\ \bibinfo {pages} {3685}
  (\bibinfo {year} {1989})}\BibitemShut {NoStop}%
\bibitem [{\citenamefont {Fagas}\ \emph {et~al.}(1999)\citenamefont {Fagas},
  \citenamefont {Kozorezov}, \citenamefont {Lambert}, \citenamefont {Wigmore},
  \citenamefont {Peacock}, \citenamefont {Poelaert},\ and\ \citenamefont {den
  Hartog}}]{Fagas:1999p3575}%
  \BibitemOpen
  \bibfield  {author} {\bibinfo {author} {\bibfnamefont {G.}~\bibnamefont
  {Fagas}}, \bibinfo {author} {\bibfnamefont {A.}~\bibnamefont {Kozorezov}},
  \bibinfo {author} {\bibfnamefont {C.}~\bibnamefont {Lambert}}, \bibinfo
  {author} {\bibfnamefont {J.}~\bibnamefont {Wigmore}}, \bibinfo {author}
  {\bibfnamefont {A.}~\bibnamefont {Peacock}}, \bibinfo {author} {\bibfnamefont
  {A.}~\bibnamefont {Poelaert}}, \ and\ \bibinfo {author} {\bibfnamefont
  {R.}~\bibnamefont {den Hartog}},\ }\href@noop {} {\bibfield  {journal}
  {\bibinfo  {journal} {Phys. Rev. B}\ }\textbf {\bibinfo {volume} {60}},\
  \bibinfo {pages} {6459} (\bibinfo {year} {1999})}\BibitemShut {NoStop}%
\bibitem [{\citenamefont {Mingo}\ and\ \citenamefont
  {Yang}(2003)}]{Mingo:2003p110}%
  \BibitemOpen
  \bibfield  {author} {\bibinfo {author} {\bibfnamefont {N.}~\bibnamefont
  {Mingo}}\ and\ \bibinfo {author} {\bibfnamefont {L.}~\bibnamefont {Yang}},\
  }\href@noop {} {\bibfield  {journal} {\bibinfo  {journal} {Phys. Rev. B}\
  }\textbf {\bibinfo {volume} {68}},\ \bibinfo {pages} {245406} (\bibinfo
  {year} {2003})}\BibitemShut {NoStop}%
\bibitem [{\citenamefont {Zhao}\ and\ \citenamefont
  {Freund}(2005)}]{Zhao:2005p024903}%
  \BibitemOpen
  \bibfield  {author} {\bibinfo {author} {\bibfnamefont {H.}~\bibnamefont
  {Zhao}}\ and\ \bibinfo {author} {\bibfnamefont {J.~B.}\ \bibnamefont
  {Freund}},\ }\href@noop {} {\bibfield  {journal} {\bibinfo  {journal} {J.
  Appl. Phys.}\ }\textbf {\bibinfo {volume} {97}},\ \bibinfo {pages} {024903}
  (\bibinfo {year} {2005})}\BibitemShut {NoStop}%
\bibitem [{\citenamefont {Wang}\ and\ \citenamefont
  {Wang}(2006)}]{Wang:2006p3815}%
  \BibitemOpen
  \bibfield  {author} {\bibinfo {author} {\bibfnamefont {J.}~\bibnamefont
  {Wang}}\ and\ \bibinfo {author} {\bibfnamefont {J.-S.}\ \bibnamefont
  {Wang}},\ }\href@noop {} {\bibfield  {journal} {\bibinfo  {journal} {Phys.
  Rev. B}\ }\textbf {\bibinfo {volume} {74}},\ \bibinfo {pages} {054303}
  (\bibinfo {year} {2006})}\BibitemShut {NoStop}%
\bibitem [{\citenamefont {Duchemin}\ and\ \citenamefont
  {Donadio}(2011)}]{Duchemin:2011p3936}%
  \BibitemOpen
  \bibfield  {author} {\bibinfo {author} {\bibfnamefont {I.}~\bibnamefont
  {Duchemin}}\ and\ \bibinfo {author} {\bibfnamefont {D.}~\bibnamefont
  {Donadio}},\ }\href@noop {} {\bibfield  {journal} {\bibinfo  {journal} {Phys.
  Rev. B}\ }\textbf {\bibinfo {volume} {84}},\ \bibinfo {pages} {115423}
  (\bibinfo {year} {2011})}\BibitemShut {NoStop}%
\bibitem [{\citenamefont {Landry}\ and\ \citenamefont
  {McGaughey}(2009)}]{Landry:2009p3275}%
  \BibitemOpen
  \bibfield  {author} {\bibinfo {author} {\bibfnamefont {E.~S.}\ \bibnamefont
  {Landry}}\ and\ \bibinfo {author} {\bibfnamefont {A.~J.~H.}\ \bibnamefont
  {McGaughey}},\ }\href@noop {} {\bibfield  {journal} {\bibinfo  {journal}
  {Phys. Rev. B}\ }\textbf {\bibinfo {volume} {80}},\ \bibinfo {pages} {165304}
  (\bibinfo {year} {2009})}\BibitemShut {NoStop}%
\bibitem [{\citenamefont {Mingo}(2006)}]{Mingo:2006p3573}%
  \BibitemOpen
  \bibfield  {author} {\bibinfo {author} {\bibfnamefont {N.}~\bibnamefont
  {Mingo}},\ }\href@noop {} {\bibfield  {journal} {\bibinfo  {journal} {Phys.
  Rev. B}\ }\textbf {\bibinfo {volume} {74}},\ \bibinfo {pages} {125402}
  (\bibinfo {year} {2006})}\BibitemShut {NoStop}%
\end{thebibliography}

%

\end{document}